\documentstyle[12pt]{article}  

\def\be{\begin{equation}}
\def\ee{\end{equation}}

\def\ba{\begin{eqnarray}}
\def\ea{\end{eqnarray}}
\def\nn{\nonumber}

\def\ov#1{\overline{#1} }
\def\ra{\rightarrow}

\begin{document}

 \title{Vector Condensate Model of 
Electroweak Interactions\footnote{Dedicated to Prof. G. Marx on his 70th birthday. }}

\author{ G. Cynolter, A. Bodor and G. P\'ocsik \\
  Institute for Theoretical Physics, \\
  E\"otv\"os Lorand University, Budapest }
  \date{}
\maketitle

\begin{abstract}

{Motivated by the fact that the Higgs is not seen, we have proposed a
version of the standard model where the scalar doublet is replaced by a vector
doublet and its neutral member forms a nonvanishing condensate.  Gauge fields
are coupled to the new vector fields B in a gauge invariant way leading to mass
terms for the gauge fields by condensation.  B-particles become massive because
of their self interactions.  Fermion and gauge field couplings are standard.
Low energy charged current phenomenology fixes the condensate.  Fermion masses
are coming from the condensation and B-particle--fermion couplings.  The
Kobayashi-Maskawa description is unchanged.  The model has a low mementum scale
of about 2 TeV.  For instance, from tree-graph unitarity at a scale of 1 TeV the
minimum mass of a charged B-particle is 369 GeV.  Such B-particles are shown to
copiously produced at high--energy linear $e^+e^-$ colliders.  The model survives
the test of oblique radiative corrections.  To each momentum scale there exists
a range of B masses where the S, T parameters are compatible with the
experiment.  For instance, at a scale of 1 TeV from the S parameter the minimum
mass of a charged (neutral) B is 200--350 GeV (400--550 GeV).}

\end{abstract}

\section{Introduction} 
 
A popular description of the symmetry breaking sector of the standard 
model is through a weakly interacting system (Higgs,$ M_H \leq$ 1 TeV). Another 
possibility is a symmetry breaking system interacting strongly with the 
longitudinal weak vector bosons. This idea has been realised in the DHT model 
[1] based on a chiral Lagrangian approach. An alternative description of the 
strongly interacting symmetry breaking system has been proposed in the BESS 
Model [2] through nonlinear realisations. 
 
Recently, top-quark condensation has also been suggested for describing 
the electroweak symmetry  breaking [3] which has resulted in several interesting 
studies (e.g. [4]). 
 
In the present note we start with the usual Lagrangian of the standard 
model of electroweak interactions, but instead of the Higgs-doublet a Y=1 
vector-doublet $B_{\mu}$ is introduced whose neutral 
component forms a condensate d. 
This creates a mechanism of dynamical symmetry breaking, and through the 
interaction of $B_{\mu}$ and the gauge fields one gets nonvanishing masses for W and 
Z, as well as a vanishing photon mass. Identifying $(-6d)^{1/2} =246 $ GeV from
the low energy charged weak current interaction yields 
the standard description of weak vector boson masses. 
 
A quartic, invariant self-coupling gives mass to $B^{0,+}$.  In
a cutoff field theory, however, the fixed value of the condensate
confines considerably the region of validity of the model [5]:
$\Lambda, m_{B^0} \leq 2.6 $ TeV for $\Lambda \geq m_{B^0}$.

Fermion mass generation by a $B^0$--condensate is possible only if we 
assume a noninvariant interaction as a start. In this case the 
usual Kobayashi--Maskawa parametrisation immediately emerges. 
 
The spin--one particle B has pair interactions with 
$ \ov{f}f, WW, ZZ, Z, \ov{B}B $  etc. 
$ \ov{f}f \ov{B}^0B^0 $ is weaker than  $ \ov{f}f H$ (Higgs), 
but both of them are proportional to $m_f$. In coupling strengths
$\ov{B}^0B^0WW (ZZ)=HHWW (ZZ), \; \ov{B}^0 B^0 Z \simeq \ov{f} f Z $. 
 
From tree--graph unitarity the allowed region of $B^+ (B^0)$ mass is estimated 
as $m_+ \geq 369 $ GeV $m_0 \geq 410$ GeV at $\Lambda=1$ TeV [6]. 
Such B's are copiously produced 
at high-energy linear $e^+e^-$ colliders [7].  
 
	As for the oblique radiative corrections, to each momentum scale there 
exist a domain of the masses of charged and neutral vector bosons where S is 
compatible with the experiments. At a scale of 1 TeV this requires vector boson 
masses of at least $m_0 \simeq $400--500 GeV , $m_+ \simeq$200--350 GeV [8].
The model survives also the test of the $\rho$ parameter [9].
For a fixed $\Lambda$  and $m_0$ the test of $\rho$  increases 
the minimum  $m_+$ coming from S. 
 
The model is outlined in Section 2 while Section 3 
contains implications of the model. 
 
\section{ The model} 
 
We replace the standard model Higgs-doublet by a Y=1 doublet of vector 
fields, 
\begin{equation} 
 B_{ \mu } = \pmatrix{ B_{ \mu }^{(+)} \cr 
 B_{ \mu }^{(0)} \cr},
\end{equation}
and assume that $B^{(0)}_{\mu}$ forms a nonvanishing condensate d, 
\begin{equation} 
 \left \langle B_{ \mu }^{(0)+}(x) B_{ \nu }^{(0)}(x) \right \rangle_0  
  = g_{ \mu \nu } d , \qquad
\left \langle B_{ \mu }^{(+)+} B_{ \nu }^{(+)} \right \rangle_0 =0
\end{equation}
$ B_{\mu}$ is coupled to itself and the SU(2) and U(1) gauge fields $A_{\mu}$
and $C_{\mu}$, 
respectively, in a gauge invariant way. In the Lagrangian of the standard model 
the H--A--C sector is replaced by $L_0(DB)-V(B)$ added
to the Lagrangian of gauge fields, where 
\ba
 L_0(DB) &=& -\frac{1}{2} \left ( D_{\mu} B_{\nu}- D_{\nu} B_{\mu} \right)^{+}
\left ( D^{\mu} B^{\nu}- D^{\nu} B^{\mu} \right), \nn \\
 D_{\mu} &=& \partial_{\mu}-\frac{1}{2}ig_j A_{j, \mu}-\frac{1}{2}ig'_j C_{\mu}, \\
 V(B) &=& \lambda (B^+_{\nu} B^{\nu})^2-\mu_0^2 B^+_{\nu} B^{\nu}, \lambda>0. \nn
\ea

Now, one can get bilinear mass terms either in the Lagrangian or in the 
equations of motion of two-point functions once the condensate (2) is assumed. 
In the present case the $W^{\pm}$ mass is determined 
by the total B-condensate, while 
the two neutral combinations are proportional to $B_{\mu}^{(+)+} B_{\nu}^{(+)}$
and  $B_{\mu}^{(0)+} B_{\nu}^{(0)}$, 
respectively. Therefore, a vanishing photon mass goes together with the 
assumption that  $ B_{\nu}^{(+)}$ does not form a condensate. 
This leads to the predictions 
\be
m_A=0, \; m_W=\frac{1}{2} g \sqrt{-6d}, \; 
m_Z=\frac{1}{2} \frac{g}{cos \theta_W} \sqrt{-6d},
\ee 
where d fixes the $B^0$--condensate. $(-d)^{1/2}$ plays the role 
of the vacuum expectation value of the Higgs field. 
We have from charged current phenomenology 
\be
d=-(6 \sqrt{2} G_F )^{-1}
\ee
Breaking the gauge symmetry by the $B^0$--condensate gives rise to a mass term 
also for $B^{(0)}_{\mu}$: 
\be
m_{B^0}^2=-10 d \lambda+ \mu_0^2.
\ee
In what follows let us assume $\mu_0=0$ , but we remark that 
for  $\mu_0^2<0$ (6) would 
give $-2 d \lambda$  since from the minimum of V, $\mu_0^2=8 \lambda d$
(case of spontaneous symmetry breaking). Since the field $B_{\mu}^{(+)}$ cannot be 
transformed out, it represents a physical field which 
gets its bare mass from the self--interactions of $B_{\nu}$,
\be
m_{B^+}^2=-8 d \lambda, \left( \frac{m_{B^+}}{m_{B^0}} \right )^2=\frac{4}{5}.
\ee

Fermions are assigned to the gauge group in the standard manner. Since 
the four-vector $\ov{\Psi}_L B_{\nu} \Psi_R $ is invariant 
under gauge transformations, the condensate 
d will generate a fermion mass term only if a noninvariant interaction is 
introduced: 
\be
 g^u_{ij}  \overline{ \Psi}_{iL} B^C_{\nu} u_{jR} B^{(0)}_{\nu} +
 g^d_{ij}  \overline{ \Psi}_{iL} B_{\nu} d_{jR} B^{(0)}_{\nu} + h.c. , \quad
\Psi_{iL}=\pmatrix{u_i \cr d_i \cr }_L, \; 
B_{\nu}^C= \pmatrix{B_{ \mu }^{(+)} \cr B_{ \mu }^{(0)} \cr}.
\ee
This leads to the Kobayashi-Maskawa description, too. A typical
 lepton or quark mass is 
\be
m_f=-4g_f d.
\ee 
For a fermion of mass $m_f$ the coupling strengths from (9) and the standard 
description are 
\begin{equation}
 g_f={3 \over \sqrt{2} } m_f G_F , \quad  
 g_f^{SM}=m_f (2 \sqrt{2} G_F)^\frac{1}{2}.
\end{equation}
 
The trilinear interactions of Z, $W^{\pm}$  and B are derived from (3) as 
\begin{eqnarray} 
 L  \left( B^{0} \right)&=&{ig \over 2cos \Theta_W } \partial^{\mu}
 B^{(0)\nu +} \left( Z_{ \mu} B_{\nu}^{(0)} - Z_{\nu} B^{(0)}_{\mu}
\right) + h.c., \nonumber \\ 
 L \left( B^+ {B^-} Z \right)&=&-(cos2\Theta_W) \cdot L
\left(B^{(0)} \rightarrow B^{(+)} \right), \\ 
 L  \left( B^{0}B^+W \right)&=&{ig \over \sqrt{2}} \Big[ \partial^{\mu} 
 B^{(+)\nu +} \left( W_{ \mu}^+ B_{\nu}^{(0)} -W_{\nu}^+
B^{(0)}_{\mu}\right) + \nonumber \\
& & + \partial^{\mu}  B^{(0)\nu +}  
\left( W_{ \mu}^- B_{\nu}^{(+)} -W_{\nu}^- B^{(+)}_{\mu}
\right) \Big] + h.c. \nonumber
\end{eqnarray}

The quartic interactions coming from (3) are self-couplings
of $ B^{+,0} $ and couplings of the type
$ \gamma \gamma B^+B^-$, $ZZ B^+ B^- $, $\gamma ZB^+ B^-$,
$\gamma W^+B^-B^0 $, $ZW^+B^-B^0 $, \linebreak  $W^-W^+B^-B^+ $, 
$ W^+ W^- B^0 \overline{B}^0 $, $ZZ B^0 \overline{B}^0 $.
For instance, the $VVB^0 \overline{B}^0 $  couplings are
\begin{eqnarray}
L=-B_{\nu}^{(0)+} B^{(0) \nu} \left( \frac{1}{2} 
g^2 W_{\mu}^-W^{+ \mu} +
 \frac{g^2}{4 cos^2 \theta_w } Z_{\mu} Z^{\mu} 
\right)+ \nonumber \\
B_{\mu}^{(0)+} B^{(0) \nu} 
\left( \frac{1}{2} g^2 W_{\nu}^-W^{+ \mu} +
 \frac{g^2}{4 cos^2 \theta_w } Z_{\nu} Z^{\mu} \right) .
\end{eqnarray}
$\ov{B}^0 B^0 Z$ is the strongest interaction ($\simeq \ov{f}{f}Z$).
$\ov{B}^0 B^0 VV$ is weaker than VVH and as strong 
as HHVV, V=W,Z. Similarly  $\ov{f}{f}H $ is stronger than 
$\ov{B}^0 B^0 \ov{f}{f}$ , while $ (\ov{B}^0 B^0)^2 $ may be 
weak or strong depending on $m_B$.

\section{Implications of the model} 
 
From precision measurements of the Z width and the form of 
$\Gamma (Z \ra  \ov{B}^0 B^0 )$
we get $ m_{B^0}\geq 43 $ GeV [5]. High energy $e^+e^-$ colliders 
provide excellent opportunities for studying B bosons. 
At planned luminosities the yield of B's is 
large in  $ e^+e^- \ra B \ov{B}, B \ov{B} Z $ 
up to near the maximum kinematically possible $m_B$'s [7]. 
The cross section of the $B^+B^-$ final state is 0.29 times that of $B^0 \ov{B}^0$  at equal 
masses and energies. From (11) we get
\begin{equation}
\sigma( e^+ e^- \rightarrow  B^0 \overline{B^0} )= g^4
\frac {(4 \sin^2 \theta_w -1)^2+1}{3072 \pi \cos^4 \theta_w } 
\frac{(s^2-4m_0^2)^{\frac{3}{2} } (s^2+3 m_0^2)} 
 { \sqrt{s} m_0^2 \left ( (s-m_Z^2+\frac{\Gamma_Z^2}{4})^2
+m_Z^2 \Gamma_Z^2 \right ) } .
\end{equation}
With incresasing $m_{B^0}$ after threshold the rise of the cross section is
slower and at $s\gg m^2_{B^0}$ $\sigma$ is proportional to $m_{B^0}^{-2}$. 

At the linear collider of $s^{1/2} = 500$ GeV ($m_{B^0}\leq 250$ GeV ) and taking the popular 
luminosity of 10fb$^{-1}$ it follows that even a high $B^0$ mass results in a large number 
of events. For instance, for  $m_{B^0}\leq $200-240 GeV we get more than 800-200 
events. At NLC (next linear collider) even higher masses can be searched for. At 
$s^{1/2} =$ 1.5 TeV and with 10 (100)fb$^{-1}$ one gets more than 200 (1000) events for 
$m_{B^0} \leq 500(700)$ GeV, and the yield is growing with decreasing $m_{B^0}$. Studying B 
production in hadron collisions is in progress. Oblique radiative corrections due 
to B-loops to the $\rho$  parameter have been calculated in ref. [9]. The contribution 
$\Delta \rho$  due to B-loops to $\rho$  is 
\be
\Delta \rho=\alpha T,
\ee
where T is one of the three parameters constrained by precision experiments. The 
analysis in Ref. 10 finds for beyond the standard model   
$\Delta \rho=-(0.09 \pm 0.25)\times 10^{-2}$
at $m_t = 130$ GeV, $m_H = m_Z$. 
 
The parameter T is defined by
\be
\alpha T= \frac{e^2}{s^2c^2 m_Z^2}( \ov{\Pi}_{ZZ}(0)-\ov{\Pi}_{WW}(0) )
\ee
with s = sin$\theta_W$, c = cos$\theta_W$, and it is 
calculated in one B-loop order. $\ov{\Pi}_{ik}$  is expressed 
by the $ g_{\mu \nu} $ terms of the vacuum polarization contributions 
$\ov{\Pi}_{ik}$   due to B-loops as 
\be
\Pi_{AA}=e^2\ov{\Pi}_{AA}, \quad \Pi_{ZZ}=\frac{e^2}{s^2c^2}\ov{\Pi}_{ZZ}, \quad 
\Pi_{WW}=\frac{e^2}{s^2}\ov{\Pi}_{WW}.
\ee
In a renormalizable theory T is finite. In the present model, however, it remains a 
function of $\Lambda$ , but the cutoff $\Lambda$ is not restricted by experimental comparison. 
 
Numerical analysis shows that for a given $\Lambda$  there is always an ($m_0=m_{B^0}, m_+=m_{B^+}$) 
region where $\Delta \rho$ is in agreement with the experimental limits.
For fixed $\Lambda$, at 
decreasing $m_0$, the $m_+$  range corresponding to the experimental  $\Delta \rho$   error bars 
shrinks. For instance for $\Lambda$ = 1 TeV and $m_0$=(100, 400, 800, 1000) GeV the 
$1 \sigma$ $ m_+$ region is (263.6--263.9, 629.7--635.6, 950--990, 1219--1450) GeV, 
respectively. In general  $\Delta \rho$  can be written in the form of 
$\Delta \rho=\frac{\Lambda^2}{m_Z^2}f \left(\frac{\Lambda^2}{m_0^2},\frac{\Lambda^2}{m_+^2} \right)$  
thus we get similar   $\Delta \rho = 0$ curves for different $\Lambda$'s by scaling the masses by a 
factor $\frac{\Lambda'}{\Lambda}$ . For higher $\Lambda$ the allowed mass region shrinks,
 for exmaple at $\Lambda$= 1 
TeV,$ m_0$ = 600 GeV : $m_+$ = 846--868 GeV;  = 1.5 TeV, $m_0$ = 900 GeV : $m_+$ = 
1275--1290 GeV;  = 5 TeV, $m_0$ = 3000 GeV : $m_+$ = 4264--4268 GeV. This has 
been checked up to $\Lambda$ =15 TeV. 

Turning to the S parameter we define [8] 
\ba
 \alpha S &=& 4 e^2 \big( \overline{ \Pi}'_{ZZ}(0) - (c^2\!-\!s^2)
\overline{\Pi}'_{ZA}(0)- s^2 c^2 \overline{\Pi}'_{AA}(0)
\big ), \nn \\
\ov{\Pi}_{ik}(0) &=& \frac{d}{dq^2} \ov{\Pi}'_{ik}(q^2)\vert_{q^2=0} .
\ea

An analysis [11] of precision experiments shows that
$S_{new}<0.09 (0.23) $ at 90 (95)\% C.L.  for
$m_H^{ref}=300 $GeV and assuming $m_t=174 $GeV (CDF value).
Requiring $S_{new} \geq 0 $, the corresponding constraints
are $S_{new} <0.38 (0.46) $ [5]. Since a Higgs of 300 GeV
is absent in the present model, its contribution, 0.063,
must be removed. In this way for the contribution of B we
have $S<0.15 (0.29)$ at 90 (95)\% C.L.  For $m_+, \  m_0
\geq 1.90 \Lambda$ this is fulfilled, in particular, $S
\rightarrow 0 $ for $\frac{\Lambda}{m_{+,0}} \rightarrow 0$ [8]. 
Since S is invariant multiplying $\Lambda, m_0, m_+$ by a 
common factor, allowed regions for scales different from 1 TeV easily follow. 
Higher $\Lambda$ attracts higher minimum masses.

The allowed regions by S are tightened by T. For example, at $\Lambda$ = 1 TeV, 
$m_0$ = 400 (600) GeV, the $m_+$ range allowed by S, T is $m_+$ = 630--636 (846--868) 
GeV. For higher $\Lambda$  the allowed $m_+$ region shrinks at the same $ m_0$. 
In general, $\Lambda$ remains unrestricted and suitable, heavy $ B^{+,0}$ provide small 
radiative corrections. $\Lambda$ can be restricted by taking into account unitartity 
requirements [6]. 
 
In the vector condensate model there exist many $BB \ra BB, VV \;BV \ra BV $  
type processes with $B=B^0, \ov{B}^0, B^{\pm}, \; V=W^{\pm}, Z $ . We consider them for longitudinally 
polarized external particles and calculate the J=0 partial-wave amplitudes, $ a_0$, 
from contact and one--particle exchange graphs. Unitarity requires $\vert Re a_0 \vert \leq 1/2$.
We have shown that the strongest lower bounds (200--400 GeV) are coming from 
B--B scatterings. Here the dominant contributions are derived from contact 
graphs. For example, in case of $ B^0 B^0 \ra  B^0 B^0 $, the contribution of the Z-exchange 
graph to the lower bound of 317 GeV ($\Lambda=1$ TeV ) is 4 GeV. 
 
One finds the best bounds in  $ B^+ B^- \ra  B^0 \ov{B}^0 $ leading to the s-wave amplitude 
\begin{equation}
a_0=- \frac{3}{16 \sqrt{10}} G_F \left( 
\frac{s^2}{2m_0m_+} -\left( \frac{m_0}{m_+} +
\frac{m_+}{m_0} \right) s 
+\frac{1}{2} m_0 m_+ \left( \frac{m_0}{m_+} +
\frac{m_+}{m_0} \right)^2 \right) .
\end{equation}
Applying the requirement of unitarity at the maximum possible energy $\Lambda$

\begin{eqnarray}
\Lambda=1.0 \hbox{TeV} &:& \quad m_0 \geq 410 \hbox{GeV}, 
\quad m_+ \geq 369 \hbox{GeV} 
\nonumber \\
\Lambda=1.5 \hbox{TeV} &:&  \quad m_0 \geq 741 \hbox{GeV}, 
\quad m_+ \geq 667 \hbox{GeV} \\
\Lambda=2.0 \hbox{TeV} &:&  \quad m_0 \geq 1091 \hbox{GeV}, 
\quad m_+ \geq 980\hbox{GeV}. 
\nonumber 
\end{eqnarray}

It follows that in this approximation the momentum 
scale cannot reach 2 TeV and the B bosons are heavy
particles. The bounds from $ B^+B^+ \rightarrow B^+B^+ $  are 
very close to (19) and they are in turn 
$ m_+ \geq $ 332 GeV, 615 GeV, 960 GeV. The above 
bounds imposed by the unitarity are similar to those
obtained from the S parameter.
 
In conclusion, the vector condensate model cannot be renormalized 
perturbatively, its scattering amplitudes contain polynomials in s, so that partial-
wave unitarity provides a maximum energy. In tree-graph approximation this is 
$\Lambda \simeq 2$ TeV. A rough interpretation of the condensate parameter d with a 
$ B^0$-propagator yields $\Lambda\leq 2.6 $ TeV. 
 
	At the same time, the B--particles must be heavy and B--masses cannot be 
far from $\Lambda$ . Indeed, for  $\Lambda \gg m_{+,0}$ the S parameter becomes too large, while the 
unitarity argument provides low masses and $\Lambda$  below $\Lambda$=1 TeV . 
 
This work is supported in part by OTKA I/7, No. 16248.

\end{document}